\documentstyle[aps,multicol,epsfig,pstricks]{revtex}

\begin{document}
\tighten
\draft
\title{Quantum Interference in Three Photon Down Conversion}
\author{Konrad Banaszek\cite{uw}
and Peter L. Knight}
\address{Optics Section, Blackett Laboratory, Imperial College,
Prince Consort Road, London SW7 2BZ, United Kingdom}
\date{October 29, 1996}
\maketitle

\begin{abstract}
We study degenerate three photon 
down conversion as a potential scheme for
generating nonclassical states of light
which exhibit clear signatures of
phase space interference. The Wigner function representing these
states contains an interference pattern manifesting quantum coherence
between distinct phase space components, and has substantial areas of
negativity. We develop an analytical description of the interference
pattern, which demonstrates how the oscillations of the Wigner
function are built up by the superposition principle. We analyze the
impact of dissipation and pump fluctuations on the visibility of the
interference pattern; the results suggest that some signatures of
quantum coherence can be observed 
in the presence of moderate amount of
noise. 
\end{abstract}

\begin{multicols}{2}
\section{Introduction}

The superposition principle is a fundamental ingredient of quantum
theory, resulting in interference phenomena not existing in classical
mechanics.
In atomic, molecular, and optical physics 
this striking feature of quantum mechanics can be studied within
several examples of simple quantum systems: a trapped ion, a diatomic
molecule and a single electromagnetic field mode in a cavity or free
space. In this context Schleich and Wheeler 
\cite{InterferenzImPhasenraum}
developed a phase space
picture of quantum interference. They demonstrated in the semiclassical
limit, that quantum mechanical transition probabilities are governed
by a set of simple rules in the phase space: a probability amplitude
is given by a sum of overlap areas, with phase factors defined by
an area caught between the states. 

Recent developments in quantum optics 
have generated significantly increased interest in the phase space
representation of quantum states, providing feasible schemes for
measuring the Wigner functions of a single light mode
\cite{VogeRiskPRA89,OregonTomography,BreiMullJOSA95},
the vibrational manifold of a
diatomic molecule \cite{DunnWalmPRL95}, 
and the motional state of a trapped atom \cite{LeibMeekPRL96},
or an atomic beam \cite{AtomicBeam}. 
These advances open up
new possibilities in experimental studies of the quantum superposition
principle, as the Wigner function provides direct insight into the
interference phenomena through its fringes and negativities,
and also completely characterizes the quantum state. 
Additionally, negativity of the Wigner function is a strong
evidence for the distinctness of 
quantum mechanics from classical
statistical theory. Consequently, it
is now possible to obtain full information on the coherence
properties of a quantum state by measuring its Wigner function,
instead of observing quantum interference only as fringes in
marginal distributions of single observables. 

Therefore,
schemes for generating quantum states with nontrivial phase space
quasidistributions,
especially those possessing substantial negativities, 
are of considerable interest. The system that
appears to
provide the most opportunities currently
is a trapped ion, whose quantum state
can be quite easily manipulated through interaction with suitably
applied laser beams \cite{TrappedAtom}. 
In the case of travelling optical fields, the range
of available interactions is far more restricted, 
and generating states with interesting phase space properties is a
nontrivial task from both theoretical and experimental points of
view. One of the states that most clearly illustrate quantum
interference is a superposition of two distinct coherent states
\cite{SchroedingerCat}, whose generation 
in microwave frequency range has
been recently reported \cite{BrunEtAlUNP96}. 
Production of these states in the optical
domain has been a subject of considerable theoretical interest.  
Though several ingenious schemes have been proposed
\cite{CatKerr,CatPar,CatNondem,CatCount}, they require
extremely precise control over the dynamics of the system, which makes
them very difficult to implement experimentally. 

In this paper we study degenerate three photon down conversion 
\cite{BrauMcLaPRA87,HillPRA90,BrauCavePRA90,%
DrobJexPRA92,TanaGantPRA92,BuzeDrobPRA93,DrobBuzePRA94,FelbSchiUNP96}
as a scheme for generating states of light that exhibit clear 
signatures of phase space interference. This generation
scheme seems to be quite attractive, since, as we will show, it is not
overly sensitive to some sources of noise. Additionally, numerous
experimental realizations of two photon down conversion for generating
squeezed light give a solid basis for studying higher order processes,
at least in principle, and developments in nonlinear optical materials
suggest it may be possible to re--examine higher--order nonlinear
quantum effects. 

Interference features of states generated in higher order down
conversion have been first noted by Braunstein and Caves
\cite{BrauCavePRA90}, who showed
oscillations in quadrature distributions and explained them as a
result of coherent overlap of two arms displayed by the $Q$ function.
The purpose of the present paper is to provide a detailed analysis of
the interference features, based on the Wigner function rather than
distributions of single observables. Compared to the $Q$ function,
discussed previously by several authors, the Wigner function carries
explicit evidence of 
quantum coherence in the form of oscillations and negative areas. 
These features are not visible in the $Q$ function, which 
describes inevitably noisy simultaneous measurement 
of the position and the momentum. 

The states generated in three photon down conversion cannot be
described using simple analytical formulae. It is thus necessary to
resort to numerical means in order to discuss their phase space
properties. However we will show that the
interference features can be understood with the help of simple
analytical calculations. These calculations will reveal the essential 
role of the superposition principle in creating the interference
pattern in the phase space. Experimental realization of
the discussed scheme along with detection of the Wigner function of the
generated field would be an explicit optical demonstration of
totally nonclassical quantum interference in the phase space. 

This paper is organized as follows. First, in Sec.~\ref{Sec:General},
we discuss some general properties of the Wigner function. In
Sec.~\ref{Sec:Computational} we present the numerical approach used to
deal with three photon down conversion. The Wigner function
representing states
generated in this process is studied in detail in
Sec.~\ref{Sec:Wigner}. In Sec.~\ref{Sec:Noise} we discuss 
briefly prospects of experimental demonstration of quantum
interference using the studied scheme. 
Finally, Sec.~\ref{Sec:Conclusions} concludes
the results. 

\section{General considerations}
\label{Sec:General}

Before we present the phase space picture of three photon down
conversion, let us first discuss in general 
how the interference pattern is built up in the phase space
by the superposition principle. Our 
initial considerations will closely follow previous
discussions of the semiclassical limit of the Wigner function
\cite{Semiclassical}. They will give us 
later a better understanding of
the interference 
we are concerned with
in three photon down 
conversion, and help us to derive an
analytical description of the interference pattern for this specific
case.  

We will start by considering a wave function of the form
\begin{equation}
\label{Eq:PsiAExpS}
\psi(q) = {\cal A}(q) \exp[i{\cal S}(q)],
\end{equation}
where ${\cal S}(q)$ is a real function defining the phase 
and ${\cal A}(q)$ is a slowly varying positive envelope. 
The Wigner function of this state is given by
(throughout this paper we put $\hbar = 1$):
\begin{eqnarray}
W_{\psi}(q,p) & = & \frac{1}{2\pi} \int \text{d} x \, 
{\cal A}(q-x/2) {\cal A}(q+x/2) 
\nonumber \\
& & \times \exp[-ipx - i{\cal S}(q-x/2) 
+i{\cal S}(q+x/2)].
\nonumber \\
& & 
\end{eqnarray}
Let us first separate the contribution from the direct neighborhood
of the point $q$. For this purpose 
we will expand the phase ${\cal S}(q)$
up to the linear term and take the 
value of the envelope at the point $q$, which gives
\begin{equation}
\label{Eq:Wdelta}
W_{\psi}(q,p) 
\approx
{\cal A}^{2}(q) \delta(p-{\cal S}'(q)) 
+ \ldots.
\end{equation}
Thus, this contribution is localized around 
the momentum ${\cal S}'(q)$, and creates 
a concentration along the 
``trajectory'' $(q,p={\cal S}'(q))$.
This result has a straightforward interpretation in
the WKB approximation of the energy eigenfunctions, 
where the phase ${\cal S}(q)$ is the classical 
action and its spatial derivative
yields the momentum. In this case, Eq.~(\ref{Eq:Wdelta}) simply states
that the Wigner 
function contains a positive component localized along the classical
trajectory \cite{Semiclassical}. 

We will now study more carefully the relation between the wave
function and the Wigner function, taking into account contributions
from other parts of the wave function, denoted symbolically by dots in
Eq.~(\ref{Eq:Wdelta}). To make the discussion more general, we will
take the wave function to be a superposition of finite number of
components defined in Eq.~(\ref{Eq:PsiAExpS}):
\begin{equation}
\label{Eq:psisum}
\psi(q) = \sum_{i} {\cal A}_{i}(q) \exp[i{\cal S}_{i} (q)]. 
\end{equation}
The Wigner function is in this case a sum of integrals
\begin{eqnarray}
W_{\psi}(q,p) & = & \frac{1}{2\pi} \sum_{ij} \int \text{d} x \, 
{\cal A}_{i}(q-x/2) {\cal A}_{j}(q+x/2) 
\nonumber \\
& & \times \exp[-ipx - i{\cal S}_{i}(q-x/2) 
+i{\cal S}_{j}(q+x/2)].
\nonumber \\
\label{Eq:Wsumint}
& & 
\end{eqnarray}
We will evaluate these integrals with the help of the
stationary phase approximation. The condition for the stationary 
points is given by the equation
\begin{equation}
{\cal S}'_{i}(q-x/2) + {\cal S}'_{j}(q+x/2) = 2p,
\end{equation}
which has a very simple geometrical intepretation. 
It shows that the contribution to
the Wigner function at the point $(q,p)$ comes from the points of the
``trajectories'' $(q_i,p_i={\cal S}'_{i}(q))$ 
and $(q_j, p_j = {\cal S}'_{j}(q))$ satisfying  
\begin{eqnarray}
(q_i + q_j)/2 & = & q \nonumber \\
\label{Eq:qiqjpipj}
(p_i + p_j)/2 & = & p,
\end{eqnarray}
i.e. $(q,p)$ is a midpoint of the line connecting 
the points $(q_i,p_i)$ and $(q_j,p_j)$. These points may lie either on
the same trajectory, i.e. $i=j$ or on a pair of different ones. In
particular, for $i=j$ we get that $q_i=q_j=q$ is always a stationary
point for $p={\cal S}_{i}'(q)$, which justifies the approximation
applied in deriving 
Eq.~(\ref{Eq:Wdelta}). At these points the second derivative of the
phase disappears. Therefore we will calculate them separately,
using the previous method. For the remaining pairs,
we expand the phases up to quadratic terms and perform 
the resulting Gaussian integrals. 
As before, we neglect variation of the envelopes, taking
their values at the stationary points. This yields an approximate form
of the Wigner function:
\begin{eqnarray}
W_{\psi}(q,p) & \approx & 
\sum_{i} {\cal A}^2_{i}(q) \delta (p- {\cal S}_{i}'(q))
\nonumber\\
& + & \sum_{ij} 
\sum_{q_i, q_j \atop {q_i + q_j = 2q \atop 
{\cal S}'_{i}(q_i) + {\cal S}'_{j}(q_j) = 2p}}
\frac{{\cal A}_{i} (q_i) 
{\cal A}_{j} (q_j)}{\sqrt{\pi i
({\cal S}''_{i} (q_i) - {\cal S}''_{j} (q_j))/2}}
\nonumber \\
\label{Eq:Wapprox}
& & \times \exp[ip(q_i-q_j) 
- i {\cal S}_i (q_i) 
+ i {\cal S}_j (q_j)],
\end{eqnarray}
where the second double sum excludes 
the case $i=j$ and $q_i = q_j = q$.

Thus, the Wigner function of the state defined in
Eq.~(\ref{Eq:psisum}) exhibits two 
main features. The first one is presence of positive humps localized
along ``trajectories'' $(q_i, p_i={\cal S}'(q_i))$. Any pair of points
on these trajectories gives rise to the interference pattern of the
Wigner function at the midpoint of the line connecting this pair. 
Let us note that the result that
the interference pattern in a given
area is generated by equidistant 
opposite pieces of the quasidistribution 
corresponds to the phase space picture of
the superposition of two coherent states 
\cite{VogeRiskPRA89,SchroedingerCat}, for
which the interference structure lies 
precisely in the center between
the interfering states. 

\section{Numerical calculations}
\label{Sec:Computational}

Numerical results presented in the following parts
of the paper are obtained using a model
of two quantized light modes: 
the signal and the pump, coupled by the
interaction Hamiltonian:
\begin{equation}
\label{Eq:OurHamiltonian}
\hat{H} = i\lambda [
\hat{b} (\hat{a}^\dagger)^3 -
\hat{b}^{\dagger} \hat{a}^3 ],
\end{equation}
where $\lambda$ is the coupling constant, and $\hat{a}$ and $\hat{b}$
are the annihilation operators of the signal and pump mode,
respectively. This Hamiltonian is very convenient for numerical
calculations, as it commutes with the 
operator $\hat{N} = 3\hat{a}^{\dagger}\hat{a} 
+ \hat{b}^{\dagger}\hat{b}$, and can be diagonalized separately in
each of the finite--dimensional
eigenspaces of $\hat{N}$. Details of
the basic numerical approach to these 
kinds of Hamiltonians can be found for example 
in Refs.\ \cite{DrobJexPRA92,TanaGantPRA92}.
In contrast, the limit of a classical, undepleted pump is
difficult to implement numerically due to singularities of the
evolution operator on the imaginary time axis \cite{BrauMcLaPRA87}. 

We assume that initially the signal mode is in the vacuum 
state $|0\rangle$, and the pump is in a coherent
state $|\beta\rangle$. After evolution of the system 
for the time $t$, which we calculate in the interaction picture, 
we obtain the reduced density
matrix of the signal field by performing
the trace over the pump mode:
\begin{equation}
\label{Eq:DefinitionRho}
\hat{\rho}(t) = \text{Tr}_{\text{pump}} [
e^{-i\hat{H} t } |0 \rangle\langle0| \otimes
|\beta\rangle\langle\beta| e^{i\hat{H} t} ].
\end{equation}
In general, $\hat{\rho}(t)$ describes a mixed state, as the
interacting modes get entangled in the course of evolution. 
This density matrix is then used to calculate the Wigner function and
other obervables of the signal mode studied further in the paper. 
In the discussions, we will make use of the analogy between a single
light mode and a harmonic oscillator, assigning the names
position and momentum to the quadratures
$\hat{q}=(\hat{a}+\hat{a}^{\dagger})/\sqrt{2}$ 
and $\hat{p}=(\hat{a}-\hat{a}^{\dagger})/\sqrt{2}i$, respectively. 

\section{Wigner function}
\label{Sec:Wigner}
We will restrict our studies to the regime of a strong pump and a
short interaction time. This regime is the most reasonable one from
experimental point of view, as 
strong pumping allows us to compensate for the 
usually weak effect of nonlinearity, and the short interaction time
gives us a chance to ignore or 
to suppress dissipation. We can gain some intuition
about the dynamics of the system by considering the
classical case; this is done in Appendix~\ref{App:Classical}. The most
important conclusion is that in the classical picture
the origin of the phase space is an
unstable fixed point, with three symmetric directions of growth, 
in a star--like formation. 

\begin{figure}
\begin{center}
\setlength{\unitlength}{1.in}
\begin{picture}(5,8.40)
\put(0.25,0.25){\epsfig{file=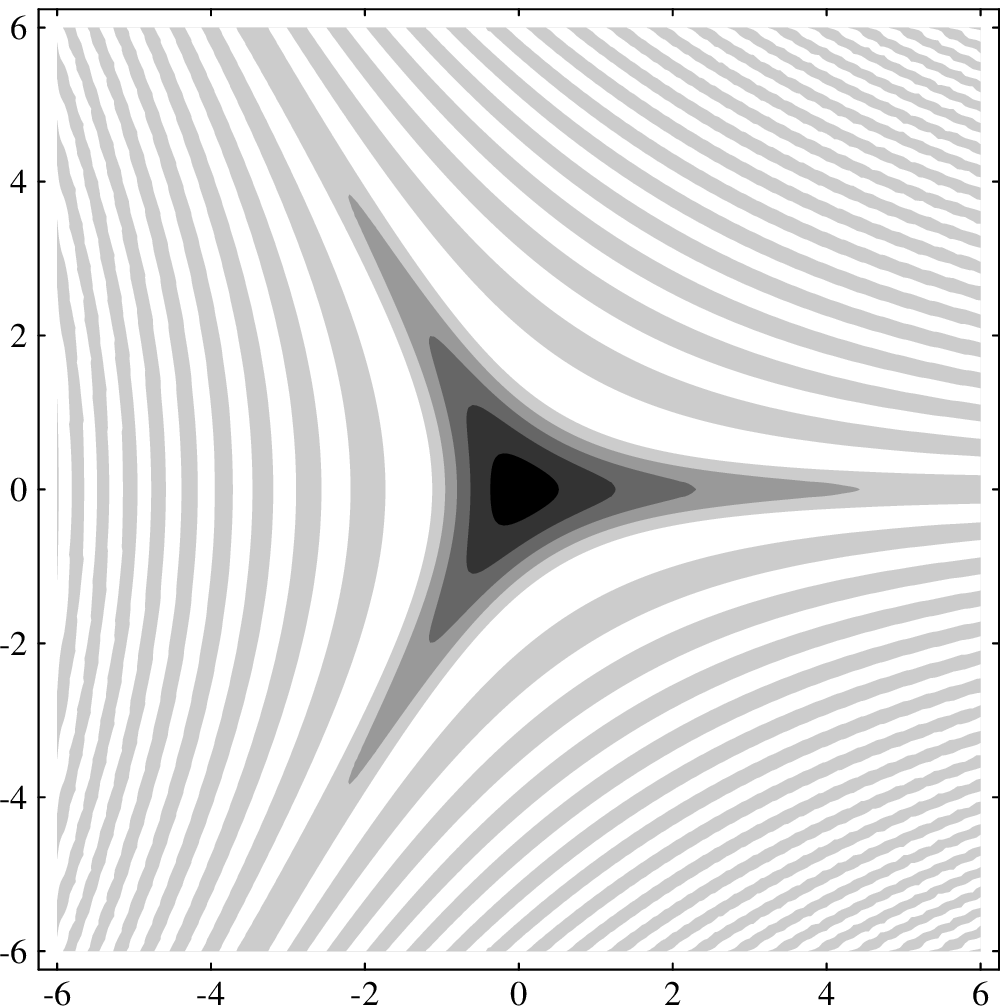}}
\put(0.25,0.25){\epsfig{file=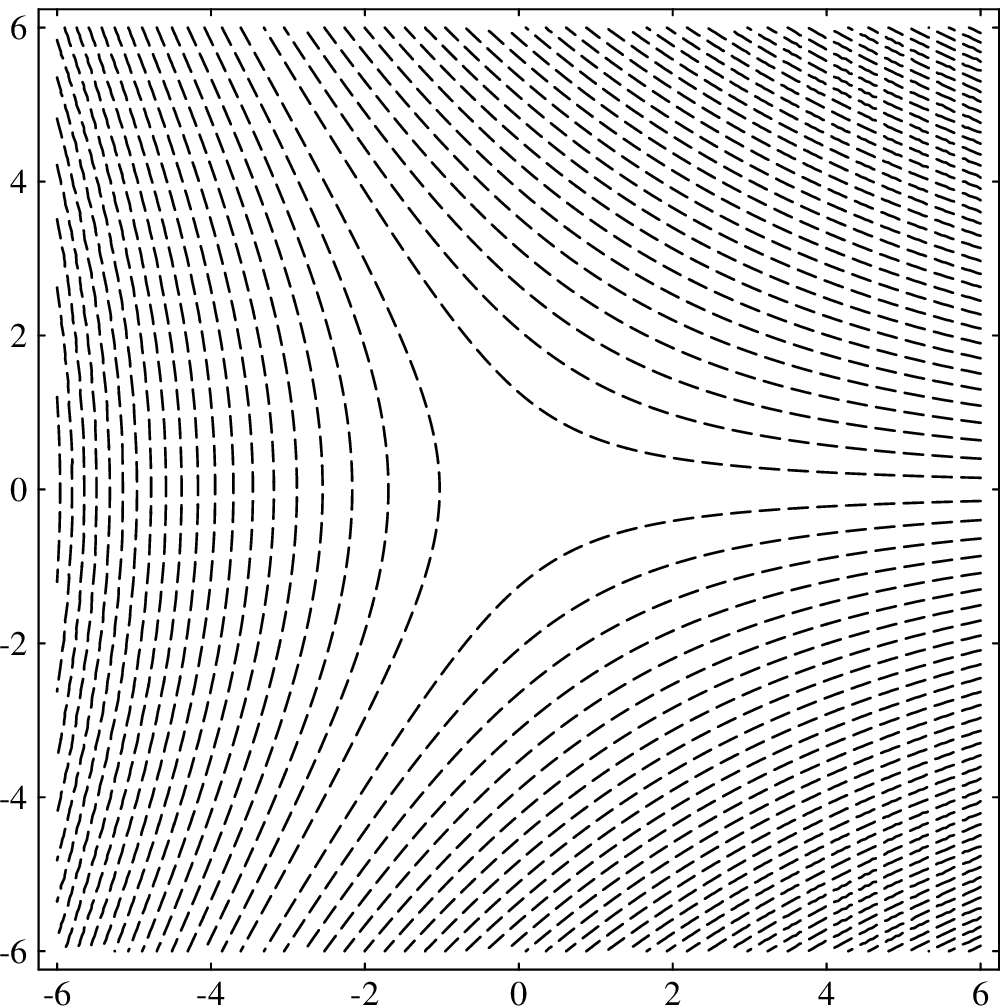}}
\put(0.00,2.25){\makebox(0,0)[lc]{\large $p$}}
\put(2.30,0.00){\makebox(0,0)[cb]{\large $q$}}
\put(0.00,8.30){\makebox(0,0)[lb]{\large $W(q,p)$}}
\put(1.60,4.80){\makebox(0,0)[cc]{\large $p$}}
\put(4.20,5.20){\makebox(0,0)[cc]{\large $q$}}
\put(5.00,8.40){\makebox(0,0)[rt]{\Large (a)}}
\put(3.70,3.85){\psframebox[linecolor=white,fillstyle=solid]{\Large (b)}}
\put(4.50,0.25){\epsfig{file=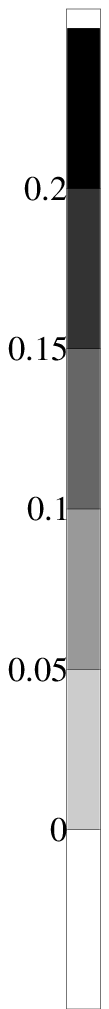}}
\put(0.00,4.25){\epsfig{file=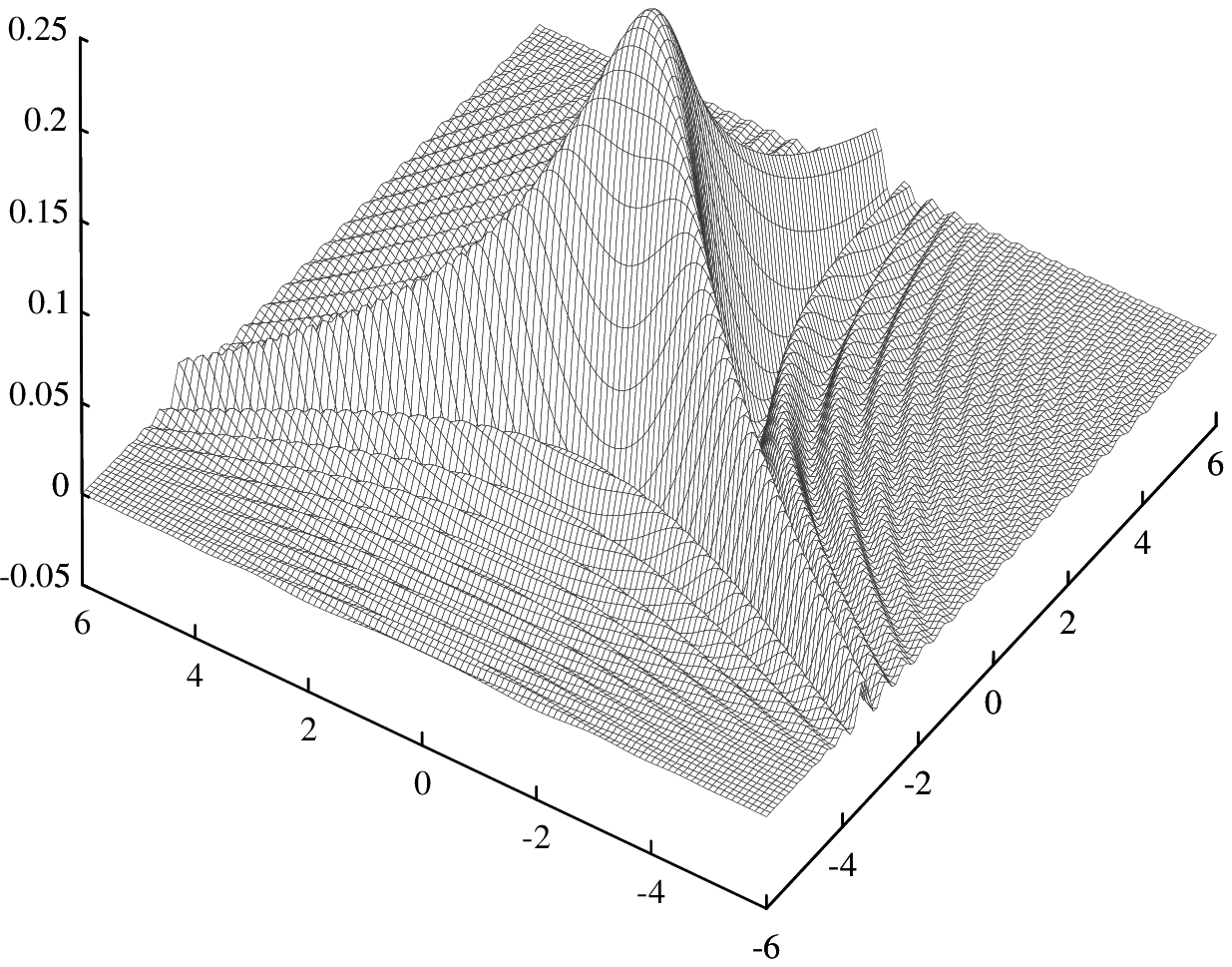}}
\end{picture}
\end{center}
\caption{The surface (a) and contour 
(b) plots of the Wigner function
representing the signal mode state 
generated for $\beta = 10$ and
$t=0.025/\lambda$. The dashed lines in the 
contour plot separate the positive
and negative regions of the 
interference term of the model Wigner
function derived in Eq.~(\protect\ref{Eq:Wintmodel}),
for comparison with the shaded plot generated from
the numerical analysis of the full model. 
\label{Fig:WignerFunction}}
\end{figure}

In Fig.~\ref{Fig:WignerFunction} we depict the Wigner function
representing the state of the signal field generated for the
parameters $\beta = 10$ and $t=0.025/\lambda$. This state is almost
pure, as $\text{Tr}[\hat{\rho}^2] = 0.92$
indicates little entanglement between
the pump and the down--converted mode. 
The three developing arms
follow the classical directions of growth from the unstable origin of
the phase space. The coherence between these components results in an
interference pattern filling the regions between the arms, consisting
of positive and negative strips. Thus, the Wigner function is
``forced'' by the superposition principle to take negative values in
order to manifest the quantum coherence of the state. 

Let us now study in more detail how the interference pattern is
generated by coherent superposition of distinct phase space
components. We will focus our attention on the three arms displayed by
the quasidistribution, neglecting the bulk of positive probability at
the origin of the phase space remaining from the initial vacuum
``source'' state. As the purity factor 
of the generated state is close to one, 
we will base our calculations on pure states. 
The relation between the wave function and the Wigner function
derived in Sec.~\ref{Sec:General} suggests that the arms can be
modelled by three components of the wave function:
\begin{equation}
\label{Eq:psi0psi1psi2}
\varphi(q) = \varphi_0(q) + \varphi_1(q) + \varphi_2 (q)
\end{equation}
with slowly varying envelopes and the position--dependent phase
factors: ${\cal S}_{0}(q) = 0$, ${\cal S}_{1}(q) = \sqrt{3}q^2/2$, 
and ${\cal S}_{2}(q) = -\sqrt{3}q^2/2$, respectively. The interference
pattern observed in the phase space is a result of the coherent
superposition of these three components. 

However, in order to calculate
quantitatively the structure of the interference pattern, we need to
know the relative phase factors between the wave functions in
Eq.~(\ref{Eq:psi0psi1psi2}). We will obtain these factors with the
help of 
the additional information that the Hamiltonian 
defined in Eq.~(\ref{Eq:OurHamiltonian}) excites or 
annihilates triplets of signal photons. Consequently, the photon
distribution of the generated state is nonzero only for Fock states
being multiples of three, as the initial state was the vacuum. 
Using this fact, we can define an
operator which performs a rotation in phase space by an 
angle $\theta$: 
\begin{equation}
\hat{U}(\theta) = \exp( -i\theta\hat{a}^{\dagger}\hat{a})
\end{equation}
and impose the relations $\varphi_1 = \hat{U}(2\pi/3)\varphi_0$ 
and $\varphi_2 = \hat{U}(4\pi/3)\varphi_0$. This choice for the phase of 
the operator $\hat{U}(\theta)$ ensures that the superposition defined
in Eq.~(\ref{Eq:psi0psi1psi2}) has the necessary property 
to generate the correct triplet photon statistics.
Let us now assume 
that $\varphi_0$ is given by a slowly varying positive 
function ${\cal A}(q)$, localized for $q>0$. 
We will not consider any specific form of the 
envelope ${\cal A}(q)$, as the main purpose of this model is to
predict the position and shape of the interference fringes. 
The other two wave
functions can be calculated with 
the help of the formula derived in
Appendix~\ref{App:Rotating}, which finally yields:
\begin{eqnarray}
\varphi_{0} (q) & = & {\cal A}(q),
\nonumber \\
\varphi_{1}(q) & = & \sqrt{2} {\cal A}(-2q) \exp ( 
\sqrt{3} i q^2/2 - i \pi/6),
\nonumber \\
\label{Eq:ThreeComponents}
\varphi_{2}(q) & = & \sqrt{2} {\cal A}(-2q) \exp ( 
-\sqrt{3} i q^2/2 + i \pi/6).
\end{eqnarray}

Given this result, we can use the approximate form of the Wigner
function in Eq.~(\ref{Eq:Wapprox}) to model the numerically calculated
Wigner function. Some problems arise from the fact that the three
components are localized along straight lines. In this case
the stationary phase approximation fails to work 
for points belonging to the same arm, 
and the Wigner 
function of each component depends substantially on the
envelope. Therefore we will 
denote them as $W_{\varphi_0}(q,p)$, $W_{\varphi_1}(q,p)$,
and $W_{\varphi_2}(q,p)$ without specifying their detailed form. 
Nevertheless,
the stationary phase approximation can be safely used to calculate 
the interference pattern between the arms, where the contributing
points belong to two distinct arms. Thus we 
represent the model Wigner function as
a sum of four components
\begin{eqnarray}
W_{\varphi}(q,p) 
& = & W_{\varphi_0}(q,p)
+ W_{\varphi_1}(q,p)
\nonumber \\
& & 
+ W_{\varphi_2}(q,p) 
+ W_{\text{int}}(q,p),
\end{eqnarray}
where the interference term $W_{\text{int}}(q,p)$ is given by
\end{multicols}
\noindent\rule{0.5\textwidth}{0.4pt}\rule{0.4pt}{\baselineskip}

\begin{equation}
\label{Eq:Wintmodel}
W_{\text{int}}(q,p) = 
\frac{4}{3^{1/4}\pi^{1/2}}
\left\{
\begin{array}{l}
\displaystyle
{\cal A}\left(-2q-\frac{2p}{\sqrt{3}}\right)
{\cal A}\left(\frac{2p}{\sqrt{3}}-2q\right)
\cos\left( \frac{p^2}{\sqrt{3}} - \sqrt{3}q^2 + \frac{\pi}{12}
\right),\makebox[2cm]{}\\
\multicolumn{1}{r}{|p|< -q} \\
\displaystyle
{\cal A}\left(2q + \frac{2p}{\sqrt{3}} \right)
{\cal A}\left(\frac{4p}{\sqrt{3}}\right)
\cos\left(\frac{2p^2}{\sqrt{3}} + 2qp - \frac{\pi}{12}
\right), \\
\multicolumn{1}{r}{p > \max\{-\sqrt{3}q,0\}} \\
\displaystyle
{\cal A}\left( 2q - \frac{2p}{\sqrt{3}} \right)
{\cal A}\left( - \frac{4p}{\sqrt{3}}\right)
\cos\left(\frac{2p^2}{\sqrt{3}} - 2qp - \frac{\pi}{12} 
\right), \\ 
\multicolumn{1}{r}{p < \min\{\sqrt{3}q,0\}}
\end{array}
\right.
\end{equation}

\hspace*{\fill}\rule[0.4pt]{0.4pt}{\baselineskip}%
\rule[\baselineskip]{0.5\textwidth}{0.4pt}
\begin{multicols}{2}
\noindent 
As the envelope ${\cal A}(q)$ 
is a positive function, the oscillations
of the interference pattern are determined by the argument of the
cosine function. The lines of constant argument are hyperbolas
with asymptotics $p =0, \pm \sqrt{3}q$. In
Fig.~\ref{Fig:WignerFunction}(b) we superpose the pattern generated
by the interference term of the model Wigner function on top of
the numerically calculated quasidistribution; the agreement between
the two is excellent. Thus, our model effectively describes the form
of the interference pattern and predicts negative areas of the Wigner
function. 

Let us emphasize that this analytical model is based exclusively on
two considerations: the position
of the interfering components in the phase
space, and the phase relations between them, which were 
derived from our study of the triplet
photon statistics for this problem.
This shows that the
interference pattern is very ``stiff'', 
i.e.\ these two considerations strictly
impose its specific form. Consequently, the interference pattern does
not change substantially as long as the crucial features of the state
remain fixed. In particular, the interaction time and the pump
amplitude have only a slight influence 
on the basic form of the interference
pattern, as they determine only the 
amount of probability density transferred
to the arms of the quasidistribution.  

\section{Consequences of phase space
interference and experimental prospects}
\label{Sec:Noise}

We will now briefly review the consequences
of these phase--space interference effects
and the prospects 
for experimental demonstration of quantum
interference using three photon down conversion. First, let us
discuss signatures of quantum coherence that can be directly observed
in the experimental data. 
An experimentally established technique for measuring the Wigner
function of a light mode is optical homodyne tomography
\cite{VogeRiskPRA89,OregonTomography,BreiMullJOSA95}. In this
method, the Wigner function is reconstructed from distributions of the
quadrature operator $\hat{x}_{\theta} = (\hat{a}e^{-i\theta} + 
\hat{a}^{\dagger} e^{i\theta})/\sqrt{2}$, measured with the help of a
balanced homodyne detector. These distributions are projections of the
Wigner function on the axis defined by the equation $q\cos\theta -
p\sin\theta = 0$. In Fig.~\ref{Fig:Quadratures} we plot the
quadrature distributions for the phase $\theta$ in the 
range $(0,\pi/6)$. Due to the symmetry of the Wigner function, other
distributions have the same form, up to the transformation
$x\rightarrow-x$. 

The fringes
appearing for $x<0$ in Fig.~\ref{Fig:Quadratures}
are a clear signature of quantum coherence between
the two arms of the quasidistribution that are projected onto the same
half--axis. We can describe the position of the fringes using the
model three--component wave function derived in
Eq.~(\ref{Eq:ThreeComponents}). For simplicity, we will consider only
the phase $\theta = 0$, for which the fringes have the best visibility
due to equal contributions from both the arms. The model quadrature
distribution  in the half--axis $x<0$ is given by
\begin{equation}
\label{Eq:ModelQuadrature}
|\varphi_1(x) + \varphi_2(x)|^2 = 8 {\cal A}^2 (-2x)
\cos^{2} (\sqrt{3}x^2/2 - \pi/6). 
\end{equation}
Analysis of this expression reveals some interesting
analogies. Expanding the argument of the cosine function around a
point $x$ yields that the ``local'' spacing between the consecutive
fringes is $\pi/\sqrt{3}x$. The same result can be obtained by
considering a superposition of two coherent states centered 
at the points $(x,\sqrt{3}x)$ and $(x,-\sqrt{3}x)$, i.e.,
where the contributing pieces of the arms are localized.
Furthermore, the argument of
the cosine function 
in Eq.~(\ref{Eq:ModelQuadrature})
is equal, up to an additive constant, to half of the
area caught between the two arms of the generated state, and the
Wigner function of the position eigenstate representing the
measurement. Thus, the quadrature distribution given in
Eq.~(\ref{Eq:ModelQuadrature}) illustrates Schleich and Wheeler's
phase space rules for calculating quantum transition probabilities
\cite{InterferenzImPhasenraum}. 

Let us now estimate the effect of dissipation and nonunit detector
efficiency on the interference pattern exhibited by the Wigner
function. For this purpose we will calculate evolution of the
generated state under the master equation:
\begin{equation}
\frac{\text{d}\hat{\rho}}{\text{d}t} = 
\frac{\gamma}{2} ( 2 \hat{a}\hat{\rho}\hat{a}^{\dagger}
- \hat{a}^{\dagger} \hat{a} \hat{\rho} - \hat{\rho}
\hat{a}^{\dagger} \hat{a}),
\end{equation}
where $\gamma$ is the damping parameter. Evolution over the
interval $\Delta t$ yields the state that is effectively measured in a
homodyne setup with imperfect detectors characterized by the quantum
efficiency $\eta = \exp(-\gamma \Delta t)$. In phase space, the
effect of dissipation is represented by coarsening of the Wigner
function by convolution with a Gaussian function \cite{LeonPaulPRA93}:
\begin{eqnarray}
W_{\eta}(q,p) & = & \frac{1-\eta}{2\pi\eta} \int \text{d}q'
\text{d}p' \, W (q',p')
\nonumber \\
& & \times \exp \left( - \frac{1-\eta}{2\eta} ((q-q')^2 + (p-p')^2)
\right).
\nonumber \\
& & 
\end{eqnarray}
This coarsening smears out entirely the very
fine details of the Wigner function, whose
characteristic length is smaller 
than $\sqrt{2\eta/(1-\eta)}$. In
Fig.~\ref{Fig:Dissipation} we plot 
the Wigner function along the
position axis as a function of $\eta$. The interference pattern
disappears faster in the area more distant from the origin of the
phase space, where the frequency of the oscillations is larger. 
Nevertheless, the first negative dip, which is the widest one, can
still be noticed even for $\eta = 0.8$. 

Current technology gives some 
optimistic figures about the possibility of
detecting the interference pattern, as virtually 100\% efficient
photodetectors are available in the range of light intensities
measured (in a different context, 
that of squeezed light) in a homodyne scheme 
\cite{PolzCarrPRL92}. However, there are
also other mechanisms of losses, such as absorption during nonlinear
interaction
and nonunit overlap of the homodyned modes,
whose importance cannot be 
estimated without reference to
a specific experimental setup.
An analysis of these would be out
of place here.

Let us finally consider the impact of pump fluctuations on the
interference pattern. We illustrate the discussion with
Fig.~\ref{Fig:PumpNoise}, depicting the state generated using a noisy
pump field modelled by a Gaussian $P$--representation
\begin{equation}
P(\beta) = \frac{1}{\pi\bar{n}} \exp 
\left( - 
\frac{|\beta - \beta_0|^2}{\bar{n}} 
\right),
\end{equation}
where $\beta_0$ is the average field amplitude and $\bar{n}$ is the
number of thermal photons. In 
discussing the effect of noise, we
have to distinguish between phase and amplitude fluctuations.
Phase fluctuations have a quite deleterious
effect, as a change in the pump
phase by $\vartheta$ is equivalent to
the rotation of the signal phase space by $\vartheta/3$. Consequently,
phase fluctuations average the signal Wigner function over a certain
phase range. The fringes are most
fragile near the arms due to
neighboring bulk of positive probability. The interference pattern in
the areas between the arms varies slowly with phase, which makes it
more robust. These properties are clearly visible in
Fig.~\ref{Fig:PumpNoise}. The effect of amplitude fluctuations is not
crucial, 
as the position of the fringes does not depend substantially on the
pump amplitude. 

\section{Conclusions}
\label{Sec:Conclusions}

We have demonstrated that degenerate three photon down conversion
generates nonclassical states of light, whose Wigner function exhibits
nontrivial interference pattern due to coherent superposition of
distinct phase space components. We have developed an analytical
description of this pattern, which precisely predicts its form. 
Let us note that the rich phase space picture of higher order down
conversion contrasts with the two photon case, where the only
signature of quantum coherence is suppression of quadrature dispersion
\cite{BuzeKnigOC91}. 

Discussion of the impact of dissipation and pump fluctuations on the
coherence properties of the generated state shows that the
interference pattern can partly be observed even in the presence of
moderate amount of noise. An important element of the studied scheme
is that the signal state is generated using a strong external pump,
which enhances the usually weak effect of $\chi^{(3)}$
nonlinearity. This allows us the optimism
to expect that three photon down conversion
is perhaps more feasible than schemes based on nonlinear
self--interaction of the signal field. 

The analytical method developed in this paper to describe the
phase space interference pattern can be applied to other cases, where
the quasidistribution is a coherent superposition of well localized
components, for example superpositions of two squeezed states
\cite{SandPRA89}, and squeezed coherent states for the SU(1,1)
group \cite{PrakAgarPRA94}. 

\section*{Acknowledgements}
This work was supported in part by the
UK Engineering and Physical Sciences
Research Council and the European Union.
K.B. thanks the European Physical
Society for support from the EPS/SOROS
Mobility Scheme. We wish to acknowledge
useful discussions with I. Jex, M. Hillery,
V. Bu\v{z}ek, and K. W\'{o}dkiewicz.

\appendix

\section{Classical dynamics of three photon down conversion}
\label{App:Classical}

The dynamics of multiphoton down conversion
under classical and quantum equations
of motion 
has been compared in detail 
by Braunstein and McLachlan \cite{BrauMcLaPRA87};
see also Ref.\ \cite{DrobBandPRA97}. 
Here, for completeness, we briefly discuss classical
trajectories for three photon down conversion
in the approximation of a constant pump. 
As the change in the pump phase is 
equivalent to the rotation of the 
signal phase space, we can assume with no loss of
generality that the pump amplitude $\beta$ 
is a real positive number. We will
now decompose the complex signal field amplitude
into its modulus $u$ and the phase $\theta$.
The classical Hamiltonian in this parametrization reads
\begin{equation}
H(u,\theta) = 2\lambda\beta u^3 \sin 3\theta,
\end{equation}
and the resulting equations of motion are
\begin{eqnarray}
\frac{\text{d}u}{\text{d}\tau} & = & 3 u^2 \cos 3\theta,
\nonumber \\
\label{Eq:ClassEqMotion}
\frac{\text{d}\theta}{\text{d}\tau} & = & 
- 3 u \sin 3 \theta,
\end{eqnarray}
where $\tau = 2 \lambda\beta t$ is the rescaled time. As the
energy of the system is conserved, trajectories of the system are
defined by the equation $u^3\cos 3\theta = \text{const}$. Thus,
trajectories are of hyperbolic--like shape, with asymptotic 
phases equal to multiples of $\pi/3$. The direction of motion can be
read out from Eqs.~(\ref{Eq:ClassEqMotion}), showing 
that the sign of the derivative $\text{d}\theta/\text{d}\tau$ 
is negative for the phases in the
intervals $(0,\pi/3)$, $(2\pi/3,\pi)$, and $(4\pi/3,5\pi/3)$, and
positive in the remaining areas. The 
resulting picture of dynamics is
presented in Fig.~\ref{Fig:ClasDyn}. It is seen that the origin of the
phase space is a three--fold unstable fixed point, with the direction
of growth $\theta = 0, 2\pi/3$, and $4\pi/3$. 

\section{Rotating the wave function in phase space}

\label{App:Rotating}

In this appendix we will calculate the rotation of a wave function
defined by a slowly varying positive function ${\cal A}(q)$
\begin{equation}
\psi_0 (q) = {\cal A}(q)
\end{equation}
around the origin of the phase space. An operator performing this
rotation is $\hat{U}(\theta) = \exp(-i\theta\hat{a}^{\dagger}\hat{a})$.
Its position representation is given by
\begin{equation}
\langle q | \hat{U}(\theta) | q' \rangle =
\frac{1}{\sqrt{\pi(1-e^{-2i\theta})}}
\exp
\left(
\frac{i}{2}
\frac{q^2 + q'^2}{\tan\theta}
-i
\frac{qq'}{\sin\theta}
\right),
\end{equation}
where the square root in the complex plane is defined 
by $\sqrt{re^{i\phi}} = \sqrt{r}e^{i\phi/2}$ 
for $r\ge 0$ and $-\pi < \phi < \pi$. The wave function rotated by an
angle $\theta$ is thus given by the integral
\begin{eqnarray}
\psi_{\theta}(q) & = &
\frac{1}{\sqrt{\pi(1-e^{-2i\theta})}} 
\nonumber \\
& & \times
\int \text{d}q'\,
{\cal A}(q')
\exp
\left(
\frac{i}{2}
\frac{q^2 + q'^2}{\tan\theta}
-i
\frac{qq'}{\sin\theta}
\right)
\end{eqnarray}
The stationary phase point for the 
exponential factor is $q'=q/\cos\theta$. We
will take the value of ${\cal A}(q')$ at this point and perform the
integral. Some care has to be taken in choosing the proper branch of
the square root function, when simplifying the final expression. 
The easiest way to avoid problems is to consider separately four
intervals of $\theta$, between $0,\pi/2,\pi, 3\pi/2$, and $2\pi$. The
final result is:  
\begin{equation}
\psi_{\theta}(q) 
= 
\sqrt{\frac{e^{i\theta}}{\cos\theta}}
{\cal A}\left(\frac{q}{\cos\theta}\right) 
\exp\left(-\frac{iq^2}{2}\tan\theta\right).
\end{equation}

\end{multicols}

\noindent
\begin{minipage}[t]{3.375in}
\begin{figure}
\begin{center}
\setlength{\unitlength}{0.675in}
\begin{picture}(5,4)
\put(0.00,3.80){\makebox(0,0)[lb]{
$\langle \delta(x - \hat{x}_{\theta})\rangle$}}
\put(3.20,0.50){\makebox(0,0)[cc]{$x$}}
\put(0.60,0.90){\makebox(0,0)[cc]{$\theta$}}
\put(0.00,0.25){\epsfig{file=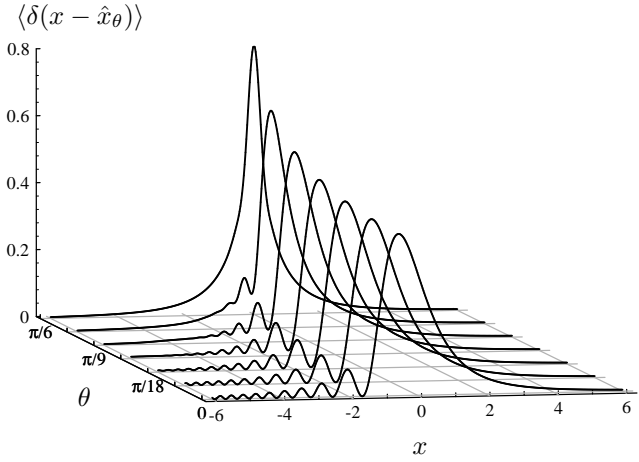,width=3.375in}}
\end{picture}
\end{center}
\caption{Quadrature distributions 
$\langle\delta(x-\hat{x}_{\theta})\rangle$
for the state plotted in
Fig.~\protect\ref{Fig:WignerFunction}. 
\label{Fig:Quadratures}}
\end{figure}

\begin{figure}
\begin{center}
\setlength{\unitlength}{0.675in}
\begin{picture}(5,6.75)
\put(0.00,6.15){\makebox(0,0)[lb]{$W_{\eta}(q,0)$}}
\put(2.85,0.65){\makebox(0,0)[cc]{$q$}}
\put(0.15,2.15){\makebox(0,0)[cc]{$\eta$}}
\put(0.00,0.00){\epsfig{file=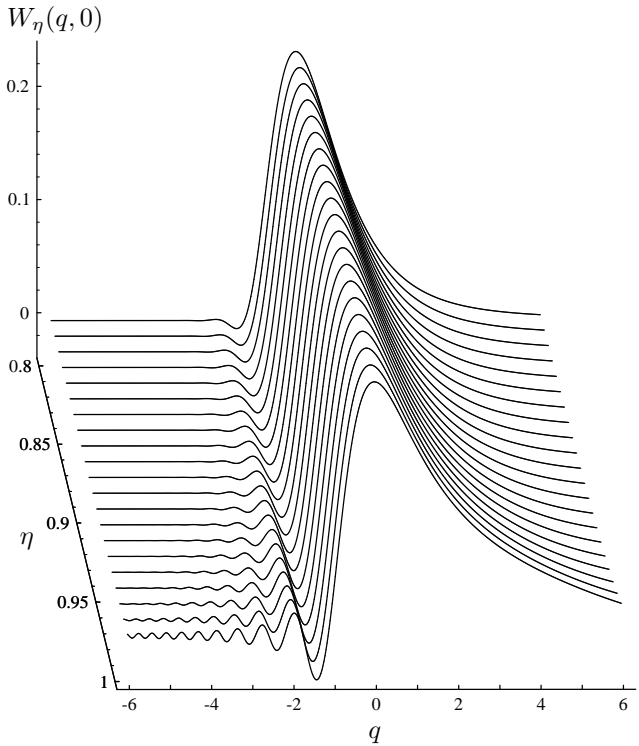,width=3.375in}}
\end{picture}
\end{center}
\caption{The Wigner function along 
the position axis $q$ 
after dissipation characterized
by the parameter 
$\eta=\exp(-\gamma\Delta t)$.
\label{Fig:Dissipation}}
\end{figure}
\end{minipage}
\hspace*{\fill}
\begin{minipage}[t]{3.375in}
\begin{figure}
\begin{center}
\epsfig{file=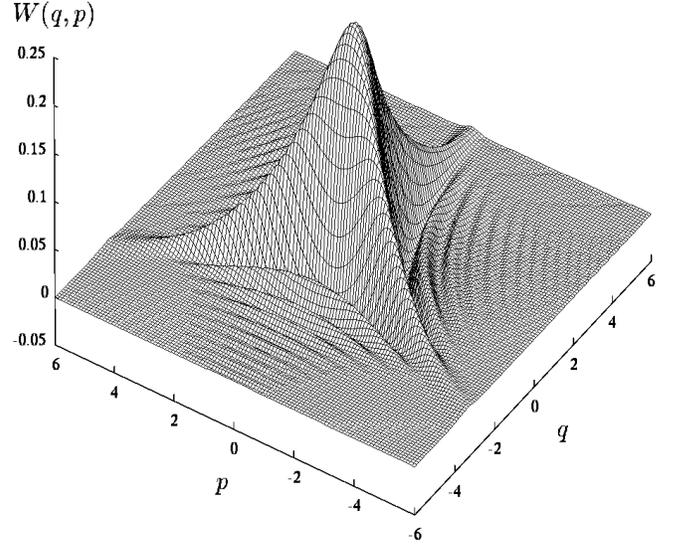,width=3.375in}
\end{center}
\caption{The Wigner function
representing the signal field state generated
using a pump with Gaussian noise, characterized by the
parameters $\beta_0=8$ and $\bar{n}=2$. The
interaction time is $t=0.025/\lambda$.
\label{Fig:PumpNoise}}
\end{figure}

\begin{figure}
\begin{center}
\epsfig{file=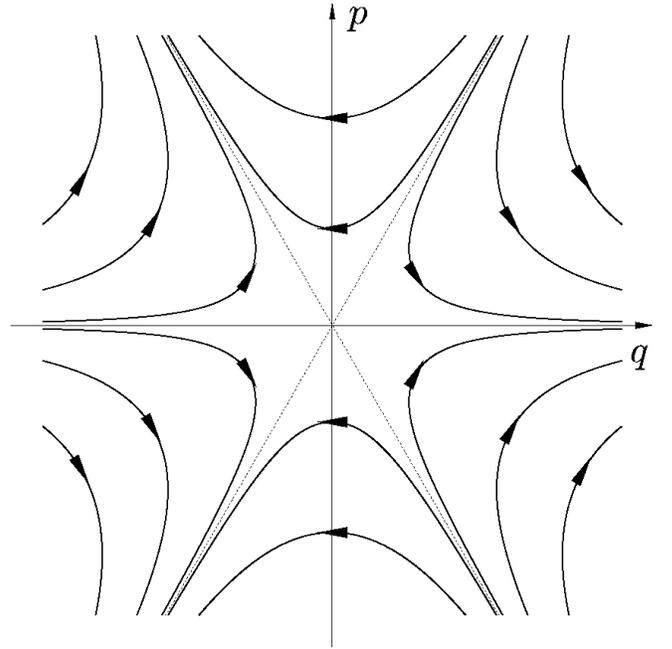,width=3.375in}
\end{center}
\caption{Classical trajectories of the
signal mode in the approximation 
of a constant pump.
\label{Fig:ClasDyn}}
\end{figure}
\end{minipage}

\end{document}